\documentclass[12pt]{article}
\usepackage{graphicx}

\def \sbar{\bar{s}}

\def \b{{\cal B}}

\def \bea{\begin{eqnarray}}
\def \beq{\begin{equation}}

\def \eea{\end{eqnarray}}
\def \eeq{\end{equation}}

\def \({\left(}
\def \){\right)}
\def \[{\left[}
\def \]{\right]}
\def \nn{\nonumber}

\def \half{\frac{1}{2}}

\def \nl{\nn\\}

\begin{document}
\rightline{TECHNION-PH-13-11}
\rightline{August 2013}
\bigskip
\centerline{\bf U-SPIN BREAKING IN CP ASYMMETRIES IN $B$ DECAYS}
\bigskip
\centerline{Michael Gronau}
\medskip
\centerline{\it Physics Department, Technion -- Israel Institute of Technology}
\centerline{\it Haifa 3200, Israel}
\bigskip
\begin{quote}
U-spin symmetry predicts equal CP rate asymmetries with opposite signs 
in pairs of $\Delta S=0$ and $\Delta S=1$ $B$ meson decays 
in which initial and final states are related by U-spin reflection. 
Of particular interest are six decay modes to final states with 
pairs of charged pions  or kaons, including $B_s \to \pi^+K^-$ and 
$B_s\to K^+K^-$ for which asymmetries have been 
reported recently by the LHCb collaboration.
After reviewing the current status of these predictions, highlighted by the 
precision of a relation between asymmetries in $B_s \to \pi^+K^-$ and 
$B^0\to K^+\pi^-$, we perform a perturbative study of  U-spin breaking corrections, 
searching for relations for combined asymmetries which hold to first order.  
No such relation is found in these six decays,
in two-body decays involving a neutral kaon,
and in three-body $B^+$ decays to charged pions and kaons.   
\end{quote}

\leftline{PACS numbers: 13.25.Ft, 13.25.Hw, 14.40.Lb, 14.40.Nd}
\bigskip

\section{Introduction\label{sec:introduction}}

Charmless hadronic $B$ decays provide valuable tests for the pattern of 
CP violation in the Cabibbo-Kobayashi-Maskawa (CKM) framework.
Methods using isospin symmetry of strong interactions have been developed 
\cite{Gronau:1990ka,Gronau:2005kz} and applied very successfully to a large amount 
of data \cite{CKMfitter,HFAG} testing the CKM framework to a high level of precision.
Flavor SU(3) relating $B$ decay amplitudes~\cite{Gronau:1994rj,Chiang:2003pm}  
is much richer than isospin alone. However, it involves symmetry-breaking effects
introducing {\it ab initio} unknown SU(3)-breaking parameters into the 
analyses~\cite{Gronau:1995hm}. These parameters have been studied using 
experimental data \cite{Gronau:2000pk,Gronau:2004ej,
Gronau:2007af,Chiang:2008vc,Imbeault:2011jz,He:2013vta}. 

A particularly useful SU(2) subgroup of flavor SU(3) is 
U-spin \cite{Meshkov:1964zz}, under which the quark pair $(d, s)$ behaves like a 
doublet while the $u$ quark and heavier quarks are singlets. 
The strangeness-conserving and  strangeness-changing parts of the effective weak 
Hamiltonian responsible for  $B$ meson decays transform like $d$ (``up") and $s$ 
(``down") components of a U-spin doublet operator. 
This and unitarity of the CKM matrix \cite{Jarlskog:1985ht}, 
${\rm Im}(V^*_{ub}V_{us}V_{cb}V^*_{cs}) =- {\rm Im}(V^*_{ub}V_{ud}V_{cb}V^*_{cd})$,
have led to the following very simple and powerful general U-spin 
prediction phrased as a theorem~\cite{Gronau:2000zy}:
{\em CP rate differences in pairs of decay processes in which both initial and final states
are obtained from each other by a U-spin reflection, $U_r: d\leftrightarrow s$, are equal
in magnitude and have opposite signs.} 
A dozen processes involving $B$ meson decays to two pseudoscalars, divided 
into half a dozen U-spin pairs obeying this theorem, were
listed in Ref. \cite{Gronau:2000zy}. Other decays involving 
one or two
vector mesons in the final state
have been discussed in the framework of U-spin in 
Refs.~\cite{Gronau:2000zy} and~\cite{Soni:2006vi}.  
Well-known examples, which have been studied extensively by experiments, are the pairs~\cite{Gronau:2000zy,Gronau:2000md,Lipkin:2005pb} 
($B^0\to K^+\pi^-, B_s \to \pi^+ K^-$) and~\cite{Fleischer:1999pa} 
($B^0\to\pi^+\pi^-, B_s\to K^+K^-$) and two pairs of three body $B^+$ decays
\cite{Bhattacharya:2013cvn}, $(B^+\to \pi^+\pi^+\pi^-, B^+\to K^+K^+K^-), 
(B^+\to\pi^+K^+K^-, B^+\to K^+\pi^+\pi^-)$.

Symmetry breaking corrections in U-spin relations between CP asymmetries have been discussed in \cite{Nagashima:2007qn} under various theoretical assumptions including factorization.  A general analysis of U-spin breaking in decays including neutral 
vector mesons in the final state has been presented in~\cite{Jung:2009pb}, with specific applications to $B$ decays into states involving a charm meson or a charmonium state. 

Typical U-spin breaking corrections estimated by $\sim (m_s - m_d)/\Lambda_{\rm QCD}$ 
or $f_K/f_\pi - 1$ are of order $20-30\%$. They may be assumed to be treated perturbatively 
in hadronic matrix elements for $B$ decays to energetic two-body final states.
Ideally, one would seek cases in which first order U-spin breaking corrections vanish or 
cancel, leaving second-order corrections which are expected to be small at a level of $5\%$.
A quantity which has been shown  recently to vanish  in the U-spin symmetry limit 
and to first order in U-spin breaking, leaving only second order U-spin breaking corections,
is the $D^0$-$\bar D^0$ mixing amplitude~\cite{Gronau:2012kq}. This property had been
shown previously to follow from a broader assumption of flavor SU(3)~\cite{Falk:2001hx}.  
Another case of SU(3) symmetry, in which first order SU(3) breaking corrections have been shown to be further suppressed by small quantities, has been studied several years ago 
in a sum rule involving decay rates for $B\to K\pi$ and $B\to K \eta^{(\prime)}$~\cite{Gronau:2006eb}.
A U-spin behavior similar to the one exhibited by $D^0$-$\bar D^0$ mixing may apply 
to certain U-spin relations among CP asymmetries in $B$ decays. The purpose of this 
Letter is to search for such relations.
 
 \section{$B^0\to K^+\pi^-, B_s\to \pi^+K^-$ and U-spin related decays\label{sec:Uspin}}
 
 In order to motivate our study we consider first the rather advanced experimental
 situation of the very early  U-spin prediction \cite{Gronau:2000zy,Gronau:2000md}  
 for the ratio of asymmetries $A_{CP}(B_s \to \pi^+K^-)/A_{CP}(B^0\to K^+\pi^-)$.
 Denoting CP rate differences by $\Delta\Gamma(B\to f) \equiv \Gamma(\bar B \to \bar f)- \Gamma(B\to f)$, the U-spin theorem quoted in the introduction 
 predicts 
 \beq\label{U}
 \Delta\Gamma(B_s\to \pi^+K^-) = - \Delta\Gamma (B^0\to K^+\pi^-)~,\
 \eeq 
 or
 \beq\label{BKpitoBsKpi}
\frac{A_{CP}(B_s \to \pi^+K^-)}{A_{CP}(B^0 \to K^+ \pi^-)}
 = -\frac{\tau(B_s)}{\tau(B^0)}
\frac{{\cal B}(B^0\to K^+\pi^-)}{{\cal B}(B_s\to \pi^+K^-)}~.
\eeq
That is, the ratio of $B_s$ and $B^0$ decay asymmetries is predicted to be negative
and equal in magnitude to the inverse ratio of corresponding decay rates. 

\begin{table}[h]
\caption{Branching fractions and direct CP asymmetries for $B^0$ and $B_s$ decays to 
pairs involving a charged pion or kaon. Data are taken from Ref. \cite{HFAG} unless 
quoted otherwise. \label{tab:BAcp}}
\begin{center}
\begin{tabular}{c c c} \hline \hline
Decay mode & Branching fraction ($10^{-6}$) & Direct CP asymmetry\\ \hline
 $B^0\to K^+\pi^-$  & $19.57^{+0.53}_{-0.52}$ & $-0.082 \pm 0.006$ \\ 
 $B_s\to \pi^+K^-$ & $5.4 \pm 0.6$  & $0.26 \pm 0.04$ \\ \hline
 $B^0 \to \pi^+\pi^-$ &   $5.10 \pm 0.19$ & $0.31 \pm 0.05~^a$ \\ 
 $B_s \to K^+ K^-$ & $24.5 \pm 1.8$ & $-0.14\pm 0.11\pm 0.03$~\cite{Aaij:2013tna} 
 \\ \hline
 $B^0 \to K^+K^-$ & $0.12 \pm 0.05$ & $-$\\ 
 $B_s \to\pi^+\pi^-$ & $0.73 \pm 0.14$ & $-$\\
 \hline \hline
\end{tabular}
\end{center}
\leftline{$^a$ World-averaged value includes $A_{CP}= 0.38 \pm 0.15 \pm 0.02$ 
from Ref.\,\cite{Aaij:2013tna}.}
\end{table}
Branching fractions and direct CP asymmetries, taken from Refs. \cite{HFAG} and 
\cite{Aaij:2013tna}, are given in Table \ref{tab:BAcp} for all $B$ and $B_s$ decays to pairs 
involving a charged pion or kaon. 
(We use the standard convention, $A_{CP}(B\to f) \equiv [\Gamma(\bar B\to \bar f)
-\Gamma(B \to f)]/[\Gamma(\bar B\to \bar f) + \Gamma(B \to f)]$.)
Several of these measurements have been  
recently improved substantially by  the LHCb collaboration \cite{Aaij:2012as,Aaij:2013iua}.
For the ratio of $B_s$ and $B^0$ lifetimes we will take the value \cite{HFAG}
$\tau(B_s)/\tau(B^0) = 0.998 \pm 0.009$.

Using the values in Table \ref{tab:BAcp} we calculate
\beq\label{eqn:DR}
-\frac{\tau(B_s)}{\tau(B^0)}
\frac{{\cal B}(B^0\to K^+\pi^-)}{{\cal B}(B_s\to \pi^+K^-)} = -3.62 \pm 0.41
\eeq
for the one side of (\ref{BKpitoBsKpi}) and 
\beq
\frac{A_{CP}(B_s \to \pi^+K^-)}{A_{CP}(B^0 \to K^+ \pi^-)} = -3.17 \pm 0.54
\eeq
for the other. 
The ratio of asymmetries is negative and larger than one around $3\sim 4$, 
consistent with the ratio 
of decay rates (\ref{eqn:DR}). Turning the argument around, one might have
used the CP asymmetry in $B^0\to K^+\pi^-$ to predict $A_{CP}(B_s \to \pi^+K^-) =
0.30 \pm 0.04$, in good agreement with the value in Table~\ref{tab:BAcp} which 
has been obtained by averaging very recent measurements by the LHCb and CDF
collaborations~\cite{Aaij:2013iua,CDF}.

The current precision of the U-spin prediction (\ref{U}) may be measured by 
the deviation of $-\Delta(B_s\to K^-\pi^+)/\Delta(B^0\to K^+\pi^-)$ from one:
\beq\label{delta}
1 +  \frac{\Delta(B_s \to \pi^+K^-)}{\Delta (B^0\to K^+\pi^-)}
= 1 + \frac{{\cal B}(B_s \to \pi^+K^-)A_{CP}(B_s \to \pi^+K^-)\tau(B^0)}
{{\cal B}(B^0\to K^+\pi^-)A_{CP}(B^0 \to K^+\pi^-)\tau(B_s)} = 0.12 \pm 0.18~.
\eeq 
U-spin breaking in decay amplitudes is enhanced by a factor four in (\ref{delta}), 
originating in the ratio of two differences of squared amplitudes for processes and
their charge-conjugates. [See Eq.\,(\ref{factor4}) in Sec.\,\ref{sec:Uspinbreak}.]
We see that the U-spin asymmetry relation (\ref{U}) is obeyed quite well. 
Current experimental errors, dominated by measurements of $\b(B_s\to \pi^+K^-)$
and $A_{CP}(B_s \to \pi^+K^-)$, allow for its violation by about $20$-$30\%$ including
this factor of four.
A resulting stringent constraint on suitably normalized U-spin breaking corrections,
at most of order several percent, will be given in Eq.\,(\ref{intrinsicU}).

The apparent success of this prediction may be accounted for by small U-spin 
breaking corrections such as occurring in an approximation based on naive 
factorization~\cite{Nagashima:2007qn}. 
A question which we will address in the next section is whether first order U-spin breaking corrections in (\ref{U}) are further suppressed or vanish in a general perturbative analysis.  
Another possibility would be to combine the rate asymmetries in (\ref{U}) with 
asymmetries in other U-spin related processes, to be discussed now, such that the combined asymmetry vanishes at first order U-spin breaking.

The other two U-spin pairs in Table \ref{tab:BAcp},  ($B^0 \to \pi^+\pi^-$, 
$B_s \to K^+K^-$), ($B^0 \to K^+K^-$, $B_s \to \pi^+\pi^-$), which require
flavor-tagging and time-dependence, involve considerably larger experimental errors. 
The U-spin relation
\beq\label{U1}
\Delta\Gamma(B_s \to K^+K^-) = - \Delta\Gamma(B^0 \to \pi^+\pi^-)
\eeq 
predicts a small negative asymmetry in $B_s \to K^+K^-$, 
\beq
A_{CP}(B_s \to K^+K^-) = 
-\frac{\tau(B_s)}{\tau(B^0)}
\frac{{\cal B}(B^0\to \pi^+\pi^-)}{{\cal B}(B_s\to K^+K^-)}\,A_{CP}(B^0 \to \pi^+\pi^-)
= -0.064 \pm 0.012~.
\eeq 
A very recent measurement reported by the LHCb collaboration~\cite{Aaij:2013tna}, 
$A_{CP}(B_s\to K^+K^-) = -0.14 \pm 0.11 \pm 0.03$, is in agreement with this prediction 
but is also consistent with zero due to a large statistical error.  It would be interesting to 
watch the change in central value with higher LHCb statistics and at the next run of 
the LHC.

Finally, the quite rare processes in the third pair, $(B^0\to K^+K^-, B_s \to \pi^+\pi^-)$, 
which are due to exchange amplitudes or final state rescattering, have been predicted to
have extremely small branching fractions~\cite{Blok:1997yj,Gronau:2012gs}.  
Asymmetry measurements in these decays which would test the U-spin prediction
\beq\label{U2}
\Delta\Gamma(B_s \to\pi^+\pi^-) = - \Delta\Gamma(B^0\to K^+K^-)
\eeq
are quite challenging. 
  
We will now use U-spin symmetry to obtain relations among the three pairs of 
processes in Table \ref{tab:BAcp}. We start by noting that the initial states
$B^0\sim \bar bd$ and  $B_s\sim \bar bs$ are members of a U-spin doublet, 
while the  $\Delta S=1$ and $\Delta S=0$ parts of the Hamiltonian, 
$H_{\rm eff}^{\Delta S=1}\sim \sbar b, -H_{\rm eff}^{\Delta S=0}\sim -\bar db$, 
transform like a U-spin doublet {\em when a minus sign is assigned to the 
$\Delta S=0$ part.}
The neutral final states involving the U-spin doublets $(\pi^-, K^-)$ and $(K^+, -\pi^+)$ 
are superpositions of U-spin singlet ($U=0$) and triplet ($U=1$) states. We 
denote $\Delta S=0$ and $\Delta S=1$ decay amplitudes into singlet and 
triplet states by $A^{d,s}_0$ and $A^{d,s}_1$, respectively. Each of these amplitudes 
consists of two terms with specific CKM factors (occasionally being referred to
as ``tree" and ``penguin" amplitudes)~\cite{Gronau:2000zy},
\bea\label{CKM}
A^d_{0,1} & = & V^*_{ub}V_{ud}A^u_{0,1} + V^*_{cb}V_{cd}A^c_{0,1}~,
\nonumber \\
A^s_{0,1} & = & V^*_{ub}V_{us}A^u_{0,1} + V^*_{cb}V_{cs}A^c_{0,1}~.
\eea

  A straight-forward U-spin decomposition gives:
 \bea\label{U-comp}
(a)~~~~~~A(B^0 \to K^+\pi^-) & = & A^s_1~,
 \nonumber\\
 A(B_s\to \pi^+K^-) & = & A^d_1~,
 \nonumber\\
(b)~~~~~~~A(B^0 \to \pi^+\pi^-) & = & \half\,A^d_1 + \half\,A^d_0~,
\nonumber\\
A(B_s \to K^+K^-) & = & \half\,A^s_1 + \half\,A^s_0~,
\nonumber\\
(c)~~~~~A(B^0 \to K^+ K^-) & = & -\half\,A^d_1 + \half\,A^d_0~,
\nonumber\\
A(B_s \to \pi^+\pi^-) & = & -\half\,A^s_1 + \half\,A^s_0~.
\eea
{\em The identical U-spin structures of amplitudes 
within each of the three pairs of processes (a) (b) and (c), each involving a U-spin 
reflection $d\leftrightarrow s$, lead to the three asymmetry relations 
(\ref{U}) (\ref{U1}) and (\ref{U2})}.
For instance, the CP rate differences $\Delta\Gamma(B^0\to K^+\pi^-)$ and 
$\Delta\Gamma(B_s \to \pi^+K^-)$ involve the same amplitude factor, 
${\rm Im}[(A^c_1)^*A^u_1]$, multiplying equal CKM factors with opposite signs,
$4{\rm Im}(V^*_{ub}V_{us}V_{cb}V^*_{cs}) =$ 
$-4{\rm Im}(V^*_{ub}V_{ud}V_{cb}V^*_{cd})$~\cite{Gronau:2000zy}.

In addition, the six amplitudes in Eqs.\,(\ref{U-comp}) are seen to obey 
two triangle relations for $\Delta S=0$ and $\Delta S=1$ transitions,
\bea\label{triangles}
A(B_s \to \pi^+K^-) - A(B^0 \to \pi^+\pi^-) + A(B^0\to K^+K^-) & = & 0~, 
\nonumber\\
A(B^0 \to K^+\pi^-) - A(B_s \to K^+K^-)  + A(B_s \to \pi^+\pi^-) & = & 0~.
\eea
Similar relations are obeyed by charged-conjugated amplitudes.
These relations  have been shown in Ref.\,\cite{Gronau:1994rj}
to hold under the broader assumption of flavor SU(3) symmetry.
We will now check the validity of these triangle relations using current 
experimental data.

Neglecting the tiny difference between $B^0$ and $B_s$ lifetimes
and omitting phase space factors, we calculate magnitudes for these six 
amplitudes and their charge-conjugates:
\beq\label{AAbar}
|A|^2 = {\cal B}(1 - A_{CP})~,~~~~~~~|\bar A|^2 ={\cal B}(1 + A_{CP})~.
\eeq
The results are summarized in Table \ref{tab:AAbar}. In the absence of asymmetry 
measurements for the rare decays $B^0\to K^+K^-$ and $B_s \to \pi^+\pi^-$ we have
assumed that these two asymmetries vanish. The magnitudes calculated in Table 
\ref{tab:AAbar} verify the closure of the two triangles (\ref{triangles}) and their 
charge-conjugates. This behavior predicted in the U-spin symmetry limit is seen to be 
independent of the values assumed for the asymmetries in  $B^0\to K^+K^-$ and 
$B_s \to \pi^+\pi^-$. 

Neglecting the amplitudes of these two rare processes relative to the 
other corresponding $\Delta S=0$ and $\Delta S=1$ amplitudes, which are almost an 
order of magnitude larger, would mean taking $A_1^{d,s} = A_0^{d,s}$. In this 
approximation the two triangles degenerate to straight lines, $A(B^0 \to \pi^+\pi^-) = 
A(B_s \to \pi^+K^-)$, $A(B_s \to K^+K^-) = A(B^0 \to K^+\pi^-)$.

%
\begin{table}
\caption{Magnitudes of amplitudes and their charge-conjugates 
for processes in Table \ref{tab:BAcp} calculated
using Eq.\,(\ref{AAbar}).\label{tab:AAbar}}
\begin{center} 
\begin{tabular}{c c c} \hline \hline
Decay mode & $|A| (10^{-3})$ & $|\bar A| (10^{-3})$ \\ \hline
 $B^0\to K^+\pi^-$  & $4.60 \pm 0.06$ & $4.24 \pm 0.06$ \\
 $B_s\to \pi^+K^-$ & $2.00 \pm 0.12$  & $2.61 \pm 0.15$ \\
 $B^0 \to \pi^+\pi^-$ &   $1.90 \pm 0.08$ & $2.56 \pm 0.07$ \\
 $B_s \to K^+ K^-$ & $4.90 \pm 0.49$ & $5.00 \pm 0.49$ \\
 $B^0 \to K^+K^-$ & $0.35 \pm 0.07$ & $0.35 \pm 0.07$\\
 $B_s \to\pi^+\pi^-$ & $0.85 \pm 0.08$ & $0.85 \pm 0.08$\\
 \hline \hline
\end{tabular}
\end{center}
\end{table}

 \section{First order U-spin breaking\label{sec:Uspinbreak}}
 
 We will now study first order U-spin breaking corrections in Eqs.\,(\ref{U-comp}),
 searching for possible relations among CP rate asymmetries which would be 
 free of such corrections.
 U-spin breaking is introduced in hadronic matrix elements by inserting a quark mass 
 term ${\cal M}_{\rm Ubrk}\propto \bar s s - \bar d d$ behaving like $U=1, U_3=0$, 
 multiplying the effective Hamiltonian which transforms as a U-spin doublet. Thus 
 the correction operator for $\Delta S=1, 0$ transitions (corresponding
 to $U_3=1/2, -1/2$) transforms as a direct product $1\otimes 1/2$ consisting of 
 $U=1/2$ and $U=3/2$ operators,
 \beq\label{MubrH}
 {\cal M}_{\rm Ubrk}\,H_{\rm eff}^{\Delta S=1,0} \propto
\pm (1,0)\otimes (\half, \pm\half) = - \sqrt{\frac{1}{3}}{\cal O}_{\pm \half}^{\half}
\pm \sqrt{\frac{2}{3}}\,{\cal O}_{\pm \half}^{\frac{3}{2}}~.
\eeq
The initial $\pm$ signs originate in the signs of $+H_{\rm eff}^{\Delta S=1}$ and 
$-H_{\rm eff}^{\Delta S=0}$ transforming as two components of a U-spin doublet. 
Upper and lower indices on operators denote values of $U$ and $U_3$, respectively.
First order corrections in (\ref{U-comp}) are given in terms of matrix elements of 
${\cal O}^{\half}$ and ${\cal O}^{\frac{3}{2}}$
for final states with $U=0,1$ and $U=1$, respectively.

 We define first order U-spin breaking corrections to decay amplitudes,  
 \bea
 \epsilon_1^{s,d} & \equiv &  
 \sqrt{\frac{1}{3}}\,\langle U=0| {\cal O}^{\half}_{\pm \half }|U=\half \rangle~,
 \nonumber\\
 \epsilon_2^{s,d} & \equiv  & 
 \sqrt{\frac{1}{3}}\,\langle U=1| {\cal O}^{\half}_{\pm \half}|U=\half \rangle~,
 \nonumber\\
 \epsilon_3^{s,d} & \equiv & 
 \sqrt{\frac{2}{3}}\, \langle U=1| {\cal O}^{\frac{3}{2}}_{\pm \half}|U=\half \rangle~.
 \eea
These corrections have CKM structures similar to (\ref{CKM})~\cite{Gronau:2000zy},
\bea\label{epsCKM}
\epsilon_i^d & = & V^*_{ub}V_{ud}\,\epsilon_i^u + 
V^*_{cb}V_{cd}\,\epsilon_i^c~,\nonumber\\
\epsilon_i^s & = & V^*_{ub}V_{us}\,\epsilon_i^u + 
V^*_{cb}V_{cs}\, \epsilon_i^c~~~~~(i=1,2,3)~.
\eea
A straight-forward U-spin decomposition including these corrections gives:
 \bea\label{U-brk}
 (a)~~~~~~A(B^0 \to K^+\pi^-) & = & A^s_1 - \epsilon_2^s - \half \epsilon_3^s~,
 \nonumber\\
 A(B_s\to \pi^+K^-) & = & A^d_1 +\epsilon_2^d + \half\epsilon_3^d~,
 \nonumber\\
(b)~~~~~~~A(B^0 \to \pi^+\pi^-) & = & \half\,A^d_1 + \half\,A^d_0 + \half \epsilon_1^d  
+ \half \epsilon_2^d - \half \epsilon_3^d~,
\nonumber\\
A(B_s \to K^+K^-) & = & \half\,A^s_1 + \half\,A^s_0 - \half \epsilon_1^s - 
\half \epsilon_2^s + \half \epsilon_3^s~,
\nonumber\\
(c)~~~~~A(B^0 \to K^+ K^-) & = & -\half\,A^d_1 + \half\,A^d_0 + \half \epsilon_1^d
- \half \epsilon_2^d + \half \epsilon_3^d~,
\nonumber\\
A(B_s \to \pi^+\pi^-) & = & -\half\,A^s_1 + \half\,A^s_0 -\half \epsilon_1^s
+ \half \epsilon_2^s - \half \epsilon_3^s~.
\eea

We note that at this order in U-spin breaking the U-spin structures of two 
amplitudes within any given pair   
are not identical as required for obtaining relations between corresponding CP 
rate asymmetries. {\em While the leading terms within each pair of amplitudes 
have the same U-spin structures and equal signs, the first order U-spin breaking 
corrections have the same structures but  opposite signs.}  
Consequently two CP rate differences for a given pair of processes now 
involve equal CKM factors with opposite signs, 
$4{\rm Im}(V^*_{ub}V_{us}V_{cb}V^*_{cs}) =$ 
$-4{\rm Im}(V^*_{ub}V_{ud}V_{cb}V^*_{cd})$, which are however multiplied by
{\em different} amplitude factors.

Denoting by $\delta$ the difference between these first order amplitude factors, 
we now have~\cite{PS}
\beq\label{del}
\Delta\Gamma(B^0\to f) + \Delta\Gamma(B_s \to U_rf) = 
4{\rm Im}(V^*_{ub}V_{us}V_{cb}V^*_{cd})\,\delta~,
\eeq
 where $U_rf$ is a final state obtained from $f$ by U-spin reflection, 
 $U_r: d \leftrightarrow s$. $\delta$ vanishes in the U-spin symmetry limit.
 Expressions of $\delta$ for the above three pairs of processes are readily 
 obtained from Eqs.\,(\ref{U-brk}):
\bea\label{delabc}
& (a) &~~\delta_a = 2{\rm Im}[A_1^{c*}(\epsilon_2^u + \half \epsilon_3^u)] - 
[c \leftrightarrow u]~,
\nonumber\\
& (b) &~~~\delta_b = \frac{1}{2}{\rm Im}[(A_1^{c*} +A_0^{c*})(\epsilon_1^u + \epsilon_2^u
-\epsilon_3^u)] - 
[c \leftrightarrow u]~,
\nonumber\\
& (c) &~~~\delta_c = \frac{1}{2}{\rm Im}[(-A_1^c{^*} +A_0^{c*})(\epsilon_1^u - \epsilon_2^u
+\epsilon_3^u)] - 
[c \leftrightarrow u]~,               
\eea
That is, all three asymmetry relations (\ref{U}), (\ref{U1}) and (\ref{U2}) obtain first 
order U-spin breaking corrections given by Eqs.\,(\ref{del}) and (\ref{delabc}). 
Furthermore, these corrections do not cancel 
in any arbitrary linear combination of the three sums of CP rate differences
(\ref{del}) because, as one can see by inspection, there exists no linear combination 
of $\delta_a, \delta_b$ and $\delta_c$ that vanishes identically.

As mentioned in Sec.\,\ref{sec:Uspin} U-spin breaking corrections in amplitudes
are enhanced by a factor four in (\ref{delta}):
\beq\label{factor4}
1 +  \frac{\Delta(B_s \to \pi^+K^-)}{\Delta (B^0\to K^+\pi^-)} =
\frac{4{\rm Im}[A_1^{c*}(\epsilon_2^u + \half \epsilon_3^u)] - 
[c \leftrightarrow u]}{{\rm Im}(A_1^{c*}A_1^u) - [c \leftrightarrow u]}~.
\eeq
Thus current measurements imply that suitably normalized  U-spin breaking
corrections are at most of order several percent,
\beq\label{intrinsicU}
\frac{{\rm Im}[A_1^{c*}(\epsilon_2^u + \half \epsilon_3^u)] - 
[c \leftrightarrow u]}{2{\rm Im}(A_1^{c*}A_1^u)} = 0.03 \pm 0.04~.
\eeq

One may also consider {\em differences} of CP rate differences for each one of 
the three U-spin pairs,
\beq\label{sigma}
\Delta\Gamma(B^0\to f) - \Delta\Gamma(B_s \to U_rf) = 
4{\rm Im}(V^*_{ub}V_{us}V_{cb}V^*_{cd})\,\sigma~.
\eeq
{\em First order U-spin breaking corrections do cancel in $\sigma$ which depends only
on U-spin invariant amplitudes},
\bea\label{Sigabc}
(a)~~~~\sigma_a & = & -2{\rm Im}(A_1^{c*}A_1^u)~,
\nl
(b)~~~~\sigma_b & = & \half{\rm Im}[(A_1^{c*} + A_0^{c*})(A_1^u + A_0^u)]~,
\nl
(c)~~~~\sigma_c & = & \half{\rm Im}[(A_1^{c*} - A_0^{c*})(A_1^u - A_0^u)]~.
\eea
These three quantities are linearly independent.
Thus in general {\em one is unable to form a linear combination of all six rate 
asymmetries which would vanish to first order in U-spin breaking.}

We checked that first order U-spin breaking cannot be avoided also
in relations between CP rate asymmetries for the following three U-reflected 
pairs of processes obeying the U-spin theorem~\cite{Gronau:2000zy}, 
$(B^0 \to K^0\pi^0, B_s \to \bar K^0\pi^0)$, $(B^0 \to K^0 \bar K^0, 
B_s \to \bar K^0 K^0)$ and $(B^+ \to K^0\pi^+, B^+\to \bar K^0 K^+)$.

In the approximation of neglecting the small amplitudes for $B^0\to K^+K^-$ and
$B_s\to \pi^+\pi^-$, which we have seen is equivalent to taking $A_1^{d,s} = A_0^{d,s}$ 
or $A_1^{c,u}=A_0^{c,u}$, one has $\sigma_a = -\sigma_b$, namely
\beq
\Delta\Gamma(B^0\to K^+\pi^-) - \Delta\Gamma(B_s \to \pi^+ K^-) = 
\Delta\Gamma(B_s \to K^+K^-) - \Delta\Gamma(B^0 \to \pi^+\pi^-)~.
\eeq
This relation holds experimentally largely because of the current large error on
$A_{CP}(B_s \to K^+K^-)$~\cite{Aaij:2013tna}. It will be interesting to watch 
the effect of improving this asymmetry measurement on the validity of this equality 
neglecting rescattering in comparison with that of (\ref{U1}) neglecting U-spin breaking.

Turning next to the amplitude triangle relations (\ref{triangles}) we observe 
that both relations are violated by purely $\Delta U = 3/2$ first order U-spin breaking 
corrections:
\bea\label{broke-triangles}
A(B_s \to \pi^+K^-) - A(B^0 \to \pi^+\pi^-) + A(B^0\to K^+K^-)   & = &
\frac{3}{2}\epsilon_3^d~, 
\nonumber\\
A(B^0 \to K^+\pi^-) - A(B_s \to K^+K^-) + A(B_s \to \pi^+\pi^-)  & = &
-\frac{3}{2}\epsilon_3^s~.
\eea

 \section{Three-body $B^+$ decays to charged pions and kaons}
 
 The LHCb collaboration reported CP 
 asymmetry measurements in all four decay modes of three-body $B^+$ decays 
 to charged pions and kaons, $B^+ \to K^+K^+K^-$, $K^+\pi^+\pi^-$, 
 $\pi^+\pi^+\pi^-$, $\pi^+K^+K^-$~\cite{Aaij:2013sfa,LHCb3P0}. These processes 
 may be divided into two pairs involving U-spin reflected final states obeying 
 relations between total CP rate asymmetries similar to (\ref{U}),
 \bea\label{U3P}
 \Delta\Gamma(B^+\to K^+K^+K^-) & = & - \Delta\Gamma(B^+\to \pi^+\pi^+\pi^-)~,
 \nonumber\\
 \Delta\Gamma(B^+\to K^+\pi^+\pi^-) & = & - \Delta\Gamma(B^+\to \pi^+K^+K^-)~.
 \eea
 These predictions have been analyzed recently in Ref.~\cite{Bhattacharya:2013cvn}
 and were found to agree reasonably well with the LHCb measurements, in particular 
 with respect to relative signs of $\Delta S=0$ and $\Delta S=1$ asymmetries and their magnitudes which currently involve sizeable errors.
  We will now study U-spin breaking corrections in (\ref{U3P}) in a manner similar to 
 our study of two-body decays in the previous two sections. 
 
 We start by observing that the initial $B^+$ state is a U-spin scalar. 
 Each one of the four final states consisting of three members of U-spin doublets can 
 be decomposed into two doublets, depending on whether the two positively charged 
 mesons combine to $U=0$ or $U=1$, and one triplet state. Only the doublets 
 contribute to the $\Delta U=1/2$ transitions in the U-spin symmetry limit.
Two amplitudes, ${\cal A}_0$ and ${\cal A}_1$, {\em defined by the U-spin of the two positively 
charged mesons}, depend also on the three meson momenta, $p_1, p_2, p_3$, defining 
a point in the Dalitz plane. 
 
Using these notations one obtains~\cite{Gronau:2003ep},
\bea\label{3K}
(d)~~~~A(B^+ \to K^+(p_1)K^+(p_2)K^-(p_3)) & = & 2{\cal A}_1^s(p_1,p_2,p_3)~,\nonumber\\
A(B^+ \to \pi^+(p_1)\pi^+(p_2)\pi^-(p_3)) & = & 2{\cal A}_1^d(p_1,p_2,p_3)~,\nonumber\\
(e)~~~~~~A(B^+ \to K^+(p_1)\pi^+(p_2)\pi^-(p_3)) & = & {\cal A}_1^s(p_1,p_2,p_3) - 
{\cal A}_0^s(p_1,p_2,p_3)~,\nonumber\\
A(B^+ \to \pi^+(p_1) K^+(p_2)K^-(p_3)) & = & {\cal A}_1^d(p_1,p_2,p_3) - 
{\cal A}_0^d(p_1,p_2,p_3)~,
\eea
where factors $1/\sqrt2$, and $1/\sqrt6$ have been absorbed in ${\cal A}_0^{d,s}$ and
${\cal A}_1^{d,s}$ which involve CKM factors as in (\ref{CKM}).
 {\em The identical U-spin structures of amplitudes within each of the two U-spin
 reflected pairs of processes (d) and (e) lead upon phase space integration to the 
 two asymmetry relations (\ref{U3P}).}
 
 In order to introduce first 
 order U-spin breaking in (\ref{3K}) we calculate additional
 contributions to decay amplitudes from the U-spin breaking  operator 
 $M_{\rm Ubrk}H_{\rm eff}$ given in (\ref{MubrH}). Corrections include matrix elements 
 of this operator for U-spin doublet final states ($|(U=1/2)_0\rangle$ and 
 $|(U=1/2)_1\rangle$, in which the two positively charged mesons combine to $U=0$ 
 and $U=1$, and for a U-spin triplet final state $|U=3/2\rangle$. Defining
 \bea
 {\cal E}_1^{s,d} & \equiv &  
 \frac{1}{\sqrt6}\,\langle (U=\half)_0| {\cal O}^{\half}_{\pm \half }|U=0 \rangle~,
 \nonumber\\
 {\cal E}_2^{s,d} & \equiv  & 
 \frac{\sqrt2}{3}\,\langle (U=\half)_1| {\cal O}^{\half}_{\pm \half}|U=0 \rangle~,
 \nonumber\\
 {\cal E}_3^{s,d} & \equiv & 
 \frac{\sqrt2}{3}\, \langle U=\frac{3}{2}| {\cal O}^{\frac{3}{2}}_{\pm \half}|U=0 \rangle~,
 \eea
 where ${\cal E}_i^{s,d}$ have CKM structures as in (\ref{epsCKM}), we calculate
 \bea\label{Ubr3K}
 (d)~~~~~A(B^+\to K^+K^+K^-) & = & 2{\cal A}_1^s - {\cal E}_2^s + {\cal E}_3^s~,
 \nonumber \\
 A(B^+\to \pi^+\pi^+\pi^-) & = & 2{\cal A}_1^d  + {\cal E}_2^d - {\cal E}_3^d~,
\nonumber \\
(e)~~~~~~~A(B^+\to K^+\pi^+\pi^-) & = & {\cal A}_1^s - {\cal A}_0^s + {\cal E}_1^s - 
\half {\cal E}_2^s - {\cal E}_3^s~,
 \nonumber \\
 A(B^+ \to \pi^+ K^+ K^-) & = & {\cal A}_1^d - {\cal A}_0^d - {\cal E}_1^d + 
 \half {\cal E}_2^d + {\cal E}_3^d~.
 \eea
 We note that, just as in two-body decays, 
 the U-spin breaking terms within each pair of three-body processes 
 have the same U-spin structures but opposite signs. Consequently the asymmetry
relations (\ref{U3P}) are violated by first order U-spin breaking corrections
in a form analogous to Eq.\,(\ref{del}). Using arguments similar to those associated with 
Eqs.\,(\ref{delabc}) and (\ref{Sigabc}) for two-body decays, we conclude that these 
corrections cannot be avoided by considering arbitrary linear combinations of these 
four asymmetries. 

Before concluding we note that while we were writing-up this work a paper 
appeared~\cite{Xu:2013dta}, in which SU(3) symmetry amplitude relations and 
SU(3) breaking corrections in these three-body decay amplitudes have been 
studied under several additional assumptions with which we do not 
completely agree. 
Among these assumptions are: (a) Two equalities, $A(B^+\to \pi^+\pi^+\pi^-)=
 A(B^+\to \pi^+K^+K^-)$ and $A(B^+\to K^+K^+K^-) = A(B^+\to K^+\pi^+\pi^-)$,
 claimed to follow from SU(3) symmetry. (b) The absence of SU(3) breaking corrections 
 in $A(B^+\to \pi^+\pi^+\pi^-)$ and $A(B^+\to K^+K^+K^-)$.  Point (a) is in clear 
 contradiction with Eqs.\,(\ref{3K}) and Bose symmetry~\cite{Gronau:2003ep} while 
 point (b) requires ${\cal E}_2^{u,c} = {\cal E}_3^{u,c}$ in Eqs.\, (\ref{Ubr3K}).

\section{Conclusion}

We have summarized the current experimental status of U-spin relations predicted 
among CP rate asymmetries in $B$ decays to two charged pseudoscalar mesons,
noticing a rather small U-spin breaking in $\Delta\Gamma(B_s \to K^-\pi^+) = - 
\Delta\Gamma(B^0\to K^+\pi^-)$. 
Introducing an $\bar s s - \bar d d$  quark mass term for U-spin breaking, we have 
performed a general analysis of first order U-spin breaking corrections in two-body 
$B^0$ and $B_s$ decays and in three-body $B^+$ decays involving charged pions 
and kaons. We have shown that these corrections cannot be made to cancel by 
a judicious choice of a linear combination of several CP rate asymmetries. 

\section*{Acknowledgments}

I am grateful to the CERN Theory Division for its kind hospitality and
wish to thank Bhubanjyoti Bhattacharya, Tim Gershon, David London 
and Jonathan Rosner for helpful comments.


\begin{thebibliography}{99}
\bibitem{PS} We neglect tiny U-spin breaking in phase space factors 
  $p_{\rm c.m.}/m^2_B$ of up to three percent, much smaller than typical first order 
  U-spin breaking in squares of amplitudes.
\end{thebibliography}
\end{document}